## Main Manuscript for

# Individuals, Crowds, and the Network Dynamics of Belief Accuracy

Charlie Pilgrim[1,a]
Joshua Becker[2,b]

[1]University of Leeds
[2]UCL School of Management

[a]c.p.pilgrim@leeds.ac.uk
[b]joshua.becker@ucl.ac.uk

**Author Contributions:** J.B. and C.P. conducted conceptual and theoretical work. C.P. prepared formal mathematical analysis. J.B. Prepared the empirical analysis. J.B. and C.P. wrote the manuscript.

**Competing Interest Statement:** The authors declare no competing interests.

**Classification:** Social Sciences / Social Sciences

**This PDF file includes:**

- Main Text
- Figures 1 to 3
- Appendix
- Figures A1 to A4

**Abstract**

Does talking to others make people more accurate or less accurate on numeric estimates such as quantitative evaluations or probabilistic forecasts? Research on peer-to-peer communication suggests that discussion between people will usually improve belief accuracy, while research on social networks suggests that error can percolate through groups and reduce accuracy. One challenge to interpreting empirical literature is that some studies measure accuracy at the group level, while others measure individual accuracy. We explain how social influence impacts belief accuracy by analyzing a formal model of opinion formation to identify the relationship between individual accuracy, group accuracy, and the network dynamics of belief formation. When opinions become more similar over time, change in individual error is always strictly better than change in group error, by a value equal to the change in variance. We show that change in group error can be decomposed into the influence network centralization, the accuracy/influence correlation ("calibration"), and the averageness/influence correlation ("herding"). Because group dynamics both theoretically and empirically lead people to become more similar over time, one might intuitively expect that the same factors which reduce group accuracy will also reduce individual accuracy. Instead, we find that individuals reliably improve under nearly all conditions, even when groups get worse. We support this analysis with data from six previously published experiments.

**Significance Statement**

Empirical research on peer-to-peer communication suggests that discussion between people will usually improve belief accuracy even as research on social networks suggests that error can percolate through groups and reduce accuracy. We explain how social influence impacts belief accuracy by analyzing a formal model of opinion formation in social networks to identify the relationship between groups and the individuals within them. We show that accuracy change can be decomposed into influence network centralization, herding (averageness/influence correlation) and calibration (accuracy/influence correlation). We find that individuals can and usually do become more accurate after communication, even when beliefs measured at the group level become less accurate. We support this argument with a reanalysis of six previously published datasets.

**Main Text**

**Introduction**

Understanding how communication between people shapes belief accuracy is a central task for social science. A range of interdisciplinary research has shown that people's beliefs are influenced by those of others, and that mere exposure to another person's estimate leads to increased similarity (1–4). Indeed, one of the most common findings in social science is that people tend to become more similar over time, through an array of mechanisms (5–9). When these processes improve individual accuracy as a result of group membership, it is known as "group to individual transfer" (10, 11).

However, prior research fails to offer any consensus on how reliable this improvement is and when this group-to-individual transfer of collective intelligence will occur. Social scientific tradition offers two compelling but competing insights. On the one hand, the popular "wisdom of crowds" principle is based on finding that group beliefs in aggregate will be more accurate than the beliefs of individual group members (4, 12–14). From this perspective, we may expect communication to improve accuracy as the wisdom of crowds is transferred from the group to the individual. On the other hand, conformity pressures have been associated with negative group dynamics (7, 15) such as "herding" (16, 17) and "groupthink" (18), suggesting that groups must be carefully managed to preserve collective intelligence (4, 19–21).

Experimental evidence also offers competing insights. Experiments on simple dyadic (one-to-one) communication suggest that social exchange is generally beneficial (1, 22–25). Some experiments suggest that individuals can benefit even while embedded in larger, more complex group contexts (10, 11, 26). However, while those studies showing benefits (10, 11, 26) enforced highly simplified social structures, a range of emergent group-level factors may cause communication to reduce belief accuracy rather than generating group-to-individual transfer (3, 27, 28). For example, communication can reduce accuracy when some group members have disproportionate influence over others, as in centralized network structures (3, 15, 27, 29).

Reconciling research on group dynamics with research on peer-to-peer communication is difficult because these two bodies of literature tend to focus on different outcomes. Research on group-to-individual transfer and advice-giving has focused specifically on individual accuracy, while research on social network theory and other emergent collective behaviors tends to focus on accuracy as measured at the group level.

We reconcile previous results and show a robust benefit of group-to-individual transfer through a formal mathematical analysis that (a) derives a clear relationship between changes in group accuracy and changes in individual accuracy (b) quantifies the effect of communication on group and individual error.

We find that error change for both groups and individuals can be decomposed into the accuracy/influence correlation (“calibration”), i.e. how much people are drawn towards accurate opinions, and the averageness/influence correlation (“herding”), i.e. how much people are drawn towards the average opinion. Alongside network centralization and the initial estimate distribution, these terms characterize collective opinion dynamics and determine change in error for both groups and the individuals within them. We find that calibration is generally good but can be too high, and that herding is generally bad, but can be too low.

Notably, individuals improve even under the conditions that reduce accuracy for groups. We show not only that individuals can often improve in conditions where groups get worse, but also that individual improvement is the most likely outcome despite the general finding that group improvement is fragile. We test key hypotheses from this model by re-analyzing six previously published datasets (3, 7, 26, 28–30).

## Formal Theoretical Analysis

**Intuition and Proof of Principle.** Under very general assumptions, individual improvement is nearly guaranteed. To see this, first note that when people become more similar over time and everyone is equally influential, people will converge toward the simple mean of pre-discussion beliefs (31). Second, note that the “crowd beats averages law” mathematically guarantees that the error of the group average is lower than the error of an average individual—this guarantee is the formal basis for the wisdom of crowds principle (14). Thus, when people converge toward the mean belief and the mean itself is unchanged, then communication will necessarily reduce average individual error.

Even outside such pristine conditions, this simple principle underlies the results presented in this paper. Suppose there is some social process that causes the average belief to become less accurate. When the change in group error is smaller than the initial crowd-wisdom advantage, then the individuals will on average become more accurate even as the group becomes less accurate.

In our analysis below, we formally define this dynamic, showing that individuals improve as long as the increase in group error is less than the change in variance of individual estimates. We then examine the robustness of this effect to variation in group behavior and communication network structure.

**Relationship between Groups and Individuals.** Following standard convention, we use the term *bias* to refer to the arithmetic difference between an estimate $X$ and the true value $\theta$, i.e. $X - \theta$ , and *error* to refer to squared bias $(X - \theta)^2$ . Note that *error* captures our intuition regarding accuracy, as it reflects the distance between an estimate and the true value, whereas *bias* reflects both distance and direction (overshoot/undershoot). At the group level, we will compare *individual error* with *group error* where the *group error* refers to the error of the mean estimate in a group (the "group estimate") and *individual error* refers to the mean of individual errors. Later, we will also consider the impact on "experts" i.e. those select individuals who are initially more accurate than average.

We find (Appendix A.1.2) that there is a fixed relationship between the change in individual error and the change in group error. Specifically, the change in individual error is equal to the change in the group error plus the change in the diversity of the opinions (as measured by the opinion variance). We indicate group error as $\bar{e}^2$ and individual error as $\overline{e^2}$, and variance as ($s_e^2$).

$$\underbrace{\Delta(\overline{e^2})}_{\text{Change in Individual Error}} = \underbrace{\Delta(\bar{e}^2)}_{\text{Change in Group Error}} + \underbrace{\Delta(s_e^2)}_{\text{Change in Opinion Diversity}} \qquad (1)$$

This relationship means that changes in average individual error are directly connected with changes in group error. Moreover, change in individual error will always be lower (better) than change in group error, as long as opinion diversity decreases ($\Delta s_e^2$ is negative) i.e. people become more similar. Thus when group error remains unchanged, the change in individual error is equal to the change in diversity and so a decrease in diversity guarantees an improvement in individual error.

Critically, individual improvement is robust to even a sizable increase in group error. In the case that opinions fully converge, individual error will only get worse if the group error increases by at least an amount equal to the variance ($s_e^2$) of the estimate distribution.

**Experts.** Although we find that individuals "on average" may improve through social influence, many people consider themselves to be "above average"—i.e., experts. An expert can always benefit in expectation from listening to the group, as long as they don't over-rely on social information. The principle of inverse-variance weighting states that the optimal self-weight will always be less than 1, i.e. even a very inaccurate group has some information about the truth, which can be used to improve the expert's estimate (Equation A10). In the case where an expert is much more accurate than the group, they can optimise their accuracy by adjusting their self-weight so that they weigh their own opinion much higher than the group. Importantly, more accurate people do in fact tend to place less weight on social information (2, 3, 28). However, even experts should always place *some* weight on the group opinion. Below, we empirically assess whether experts are likely to perform well in a group context.

**Network Model with Relaxed Assumptions.** Our analysis so far provides a guaranteed, parameter-free, and assumption-free relationship between individual and group error that holds under all circumstances. In order to study the effects of network structure including variation in influence and self-confidence, we further study a model of belief formation in networks (31).

In this model (31), people start with an independent estimate and are then influenced by other individuals simultaneously (i.e. in discrete time steps) revising their beliefs to adopt a weighted average of their own belief and peer beliefs. The weight that each person places on each other person and their own belief is fixed, and does not change over time. This model is a flexible model that allows someone to adopt the average of their friends, give all attention to one person, give different people different weights, or even completely ignore peer information.

The pattern of influence can be represented as a weighted directed adjacency matrix, i.e. a social network. Each entry in this matrix represents a network tie indicating the amount of weight a person places on their own belief (the diagonal) and others' beliefs (off-diagonal). Each person's incoming network ties including their self-weight add up to 1.

After *k* time steps, the group opinion is equal to a weighted mean of initial independent opinions. Each individual's influence on the final group opinion is described by a $k$-step influence vector, $v$, where $v_i$ indicates the amount of weight given to person $i$. The long-term influence vector for infinite time-steps is found as the leading eigenvector of the influence matrix (31). The $k$-step influence vector is the normalized row (out-weight) sums of the influence matrix raised to the $k$th power (Appendix A.3.3).

The long-term influence is also known as "centrality" in social network theory, where each individual's centrality score reflects their overall contribution to group beliefs. The extent to which that influence/centrality varies between people, versus being equally distributed, is known as the "centralization" of the social network. There are many different ways to measure centralization, but a common finding is that empirical social networks are highly centralized, meaning that some people are very central while most people are relatively peripheral (32).

Critically, experimental research has shown that network centralization is a key moderator of the effect of social influence on numeric estimate accuracy (3, 27).

We measure network centralisation as the coefficient of variation (mean-normalized variance) of the influence vector, $c_v$ which emerges as a natural measure in our analysis (Appendix A.3.4). When $c_v$ is high, some people are more influential than others and the network is said to be "centralized." In such cases, groups are influenced by the opinions of just a few central individuals and therefore reflect the “wisdom of the few” rather than the wisdom of the crowd. When $c_v$ is zero there is no variation in influence, everyone is equally influential, and the network is said to be "decentralized."

Importantly, we find that the change in group opinion after $k$ time steps (standardised by initial opinion variance) can be predicted precisely as a simple product of only two variables: i) $k$-step influence centrality measured by the coefficient of variation, ii) the correlation between influence and initial opinions (Appendix A3.4),

$$\frac{\Delta \bar{e}}{s_e} = c_v r(v, e). \tag{2}$$

Furthermore, we can use this relationship to directly calculate changes in group error and individual error (Equations A34 and A35).

**Decomposition of Group Dynamics into Calibration and Herding.** Previous experimental research (3, 28) has suggested that the correlation between influence and error (i.e. whether more accurate people are more influential) is an important determination of group dynamics. We refer to this correlation as “calibration,” following prior work (28).

Remarkably, we find that change in error can be decomposed into two distinct components, (1) the influence/accuracy correlation $-r(v, e^2)$ that we term *calibration*, and an influence/averageness correlation $-r(v, d^2)$ which we term *herding*. Here, $e^2$ indicates an individual’s error, and $d^2$ indicates an individual's deviance from the mean. A person with a small $d^2$ has a relatively average opinion and a person with a large $d^2$ has an opinion far from the group average.

This decomposition suggests that collective opinion dynamics are determined by two competing influences: the influence of accurate members, and the influence of the group average.

To express this finding, we first define a normalised measurement of the final group bias as a fraction of the initial group bias,

$$\beta = \frac{\bar{e}_k}{\bar{e}}. \tag{3}$$

Because we measure final group bias β as a percentage of initial bias, this metric reflects the change in group error as a result of communication. The change in crowd error is given by $\Delta\bar{e}^2 = \bar{e}^2(\beta^2 - 1)$, and the change in individual error is given by $\Delta\overline{e^2} = \bar{e}^2(\beta^2 - 1) + \Delta s_e^2$.

We show (Appendix A.3.10) that β and thus group accuracy is governed by the relative balance of herding and calibration. Specifically, we find that:

$$\beta = \frac{c_v}{2\bar{e}^2}\left(s_{d^2}H - s_{e^2}C\right) + 1, \tag{4}$$

where $C$ indicates calibration measured as the negative influence/error correlation $-r(v,e^2)$; $H$ indicates herding measured as the negative influence/averageness correlation $-r(d,e^2)$; $s_{d^2}$ and $s_{e^2}$ indicate the sample standard deviation of deviance $d^2$ and error $e^2$.

**Interpreting Beta.** When $\beta = 1$, this value indicates that group bias (and thus group error) is unchanged. When $\beta = 0$, the group has changed such that post-communication group error is zero. A value of $|\beta| > 1$ indicates that group error increased, while $-1 < \beta < 1$ indicates that group error decreased.

To fully interpret the dynamics represented by β, it is important to note that a group moving "towards" truth can move "too much" and thus "overshoot" the truth. When $\beta > 1$, that means that bias has increased but not changed direction—the group moved away from truth. When $\beta < -1$, the error has increased—since $|\beta|>1$—but it has flipped direction. This value can only occur if the group average moves towards truth, overshoots it, and then keeps going enough that it ends further from the truth than where it started.

In the next section, we interpret this equation and detail the effects of calibration and herding on group accuracy, showing that group accuracy is fragile even as individual accuracy is robust.

**The Effect of Calibration and Herding on Accuracy.** Equation 4 reflects the relative (im)balance between calibration and herding. When the difference between these two terms is zero i.e. when they are balanced, then $\beta = 1$ which indicates that final error is equal to initial error and the group estimate has not changed at all. Note, however, that individuals still improve in this case. This outcome is shown in Figure 1 as the blue color (improvement) for individual error (bottom row) along the $y = x$ diagonal of each panel where calibration and herding are equal.

When the herding term is greater than the calibration term, the resulting value in parenthesis in Equation 4 is positive, and $\beta > 1$, indicating that group error increases. Again, however, individual error will still improve in many of these cases. This outcome is shown in Figure 1 as the lower right corner of each heatmap panel, where $x > y$ i.e. herding is greater than calibration.

Finally, when the calibration term is greater than the herding term, then $\beta < 1$, as in the upper left corner of the heatmaps in Figure 1. In many of these cases, group error will improve, which is what we might intuitively expect when the most accurate people are also the most influential. However, when calibration is much stronger than herding, groups may move towards and then "overshoot" the truth (see discussion above). Intuitively, this can occur because the most accurate person in a group may still be less accurate than the group average—a core principle of collective intelligence. Thus extreme calibration may paradoxically reduce accuracy by generating the wisdom of the few rather than the wisdom of the crowd.

Taken together, this analysis supports two core conclusions. First, group accuracy is determined by the relative (im)balance between calibration and herding, with these values together determining change in error. Second, individual accuracy is remarkably robust, and improves under a wide range of conditions that harm group accuracy.

**Empirical Analysis**

We now test whether these dynamics are consistent with empirical results by re-analysing data from six previously published papers (3, 7, 26, 28–30). These experimental trials vary on two key dimensions, social network structure and communication modality. Three papers (3, 7, 30) used a modality in which participants observed numeric information about each other's estimates, two papers (26, 29) provided a computer-based text chat interface, and one paper (28) facilitated direct face-to-face discussion. Four of the papers (3, 7, 26, 30) also included an independent control group, in which individuals provided multiple revised estimates over time but without social exchange.

Regarding network structure, all experiments examined decentralised networks where everyone was observed by the same number of peers, and one paper (3) also included a highly centralised network. Prior evidence (29, 33) also suggests that discussion generates emergent centralization on top of explicitly controlled structure, since some people are talkative or persuasive.

We refer to a single experimental "trial" as a single group completing a single estimation task. We analyze responses by participants who complete both a pre-communication and post-communication estimate. For trials with multiple rounds, we examine the first and final estimates only.

**General Fit to Theoretical Model.** We first examine the empirical relationship between group error and individual error shown in Equation 1 by measuring the difference in the variance-normalised changes of individual error and group error.

Equation 1 states that if diversity decreases over time (i.e. if people become more similar) then the asymptotic difference is 1, and that this value will fall between 0 and 1 for finite-time (empirical) data. Out of 2,236 trials, we find that 1% show no change in variance, 9% show an increase in variance, and 90% show a decrease in variance.

As predicted by Equation 1 and shown in Figure 2, those trials showing a decrease in variance (blue points) also show a change in individual error that is lower (better) than the change in group error. In contrast, those trials showing an increase in variance (Fig. 2, yellow points) show a change in individual error that is greater (worse) than the change in group error. For all trials with a decrease in variance, the difference between change in individual error and change in group error fell between 0 and 1 units of variance, as required by Equation 1 and shown in Figure 2.

To empirically test this relationship, we note that rearranging the terms of Equation 1 yields a standard regression equation, $Y = B_0 + B_1 X$ , where $Y$ is the change in normalised group error and $X$ is the mean change in normalised individual error. Because variance converges toward zero as people become more similar, the theoretical analysis predicts asymptotically that $-1 \leq B_0 \leq 0$, and $B_1 = 1$. Visually, this relationship is shown in Figure 2 by the fact that the points cluster between the lines $Y = X$ and $Y = X - 1$.

We fit a standard linear regression to those trials with a decrease in variance, which yields a slope $B_1 = 0.998$ (95% conf. interval [0.996, 1.00]) and a mean difference represented as the intercept of the regression $-B_0=0.639$ (95% conf. interval [0.630,0.648]). Thus, empirical data shows a close fit to Equation 1.

**Effect of Social Influence on Accuracy.** While the data so far shows a strong fit to theoretical expectations, the basic question remains—are individuals actually getting more accurate? It is necessary to measure this outcome specifically because it is possible for the relationship described above to hold even as individuals get worse, if groups themselves become sufficiently worse.

To address the question of whether social influence robustly benefits individuals in the group, we directly measure the probability that an individual will become more accurate after information exchange, conditional on whether the group as a whole gets more accurate.

We consider two metrics, (1) the probability of improvement conditional on any revision at all, and (2) the probability of improving or staying the same. The first metric is informative to a person who is considering revising their belief after exposure to peer beliefs. The second metric is informative to a person or manager who is considering the risk associated with encouraging communication among team members.

Figure 3 (left) shows that social influence either helps or at least does not harm individuals in groups: across all communication modalities, and all group outcomes, individuals are significantly more likely to improve or stay the same than they are to get less accurate. The center panel shows that none of the observed conditions pose a systematic risk to individuals who revise their estimate, while most conditions show a benefit.

For all conditions, individuals who revised were more likely to improve than get worse if the group also got better. For decentralised numeric exchange and unstructured discussion, individuals were likely to improve even when the groups got worse.

Only the centralised numeric exchange condition did not lead to systematic improvement when the group as a whole got less accurate. However, even this “worst case scenario” posed no systematic risk: these outcomes were indistinguishable from both chance and from the no-influence control condition.

**Experts.** These results show overall that communication usually improves individual accuracy and, in the worst-case scenario, poses no systematic risk. Thus from an organisational or managerial perspective, encouraging communication among group members may seem like a good idea. However, from the individual perspective, there may yet be some risk: what if a person believes themselves to be unusually accurate in the first place?

To assess the effect of social influence on highly skilled contributors, we measure whether social influence improved outcomes for the most accurate versus the least accurate contributors. Critically, we don’t simply measure outcomes for those with the most initially accurate estimates—this analysis would be expected to show reduced accuracy for experts simply due to the probabilistic effect of regression to the mean. Instead, for each person for each task, we measure the error quartile *for all other completed estimates by that person*, and calculate the average across all completed tasks. The resulting value is a socially-relevant metric comparable to quartile that is not subject to regression to the mean, since the explanatory variable (initial accuracy) is being measured from a different dataset than the outcome variable (change in accuracy). We then round the resulting value to the nearest integer in order to allow us to measure average outcomes for each quartile bin.

Figure 3 (right) shows the effect of communication as a function of individual accuracy. First, we find that even the most accurate people are significantly likely to improve. Thus we can reject the alternative hypothesis that communication helps most people but hurts experts. Second, we find that individuals across accuracy levels were no more or less likely to improve. Broadly speaking, we find no evidence that the benefit of social influence varies based on skill. Notably, this result requires some careful interpretation, which we consider further in the discussion.

## Discussion

We find that the accuracy of individuals in groups robustly improves following social exchange: individuals nearly always improve, even when groups get worse.

We show that a consequence of the "crowd beats averages law" is that changes in the average accuracy of individuals are guaranteed to be more favourable than changes in the group accuracy. This finding is general, and applies to any form of dynamics whereby opinions become more similar, which is usually the case during human social exchange (1–4).

Our theoretical analysis of belief-updating in networks shows that the dynamics of belief accuracy depend on just three system variables: i) the initial group error; ii) the influence centralisation; iii) the correlation between error and influence. Notably, changes in accuracy do not depend on the particular structure of the opinion or influence distribution, beyond those three measures. Our analysis aligns with previous experimental research and provides a formal theoretical basis for the properties they identified as important (2, 3, 28).

Our empirical analysis is limited, especially with regard to experts who may have higher relative accuracy than is captured by our low-resolution quartile analysis. Notably, though, prior evidence suggests that people with true expertise are likely to give less attention to social influence (2, 3, 28, 34), in which case the risk is limited to people who do not recognize their own accuracy.

Another major limitation of our empirical analysis is that we don't measure herding or calibration. This analysis is not feasible with our present data because we cannot measure a person's influence. While one of the datasets explicitly created centralized individuals (3), the centralization in all other datasets varied endogenously (7, 26, 28–30). For example, in decentralized numeric exchange trials, the only source of influence is stubbornness—how much or little a person revises—which cannot be estimated in this data.

Our analysis highlights the fragility of collective accuracy and supports the argument that communication should be carefully managed for group decisions (3, 27, 35). However, where groups converse before making separate individual decisions, we show that the "group to individual transfer" (10, 11) is robust.

**Materials and Methods**

Replication data and code for theoretical figures, empirical analysis, and empirical figures are available at https://github.com/chasmani/Individuals-Crowds-Network-Dynamics . This code also includes numerical simulations of the DeGroot model that verify the mathematical results.

**Acknowledgements**

We are grateful to the authors of prior research for making their datasets available and upholding norms of open science.

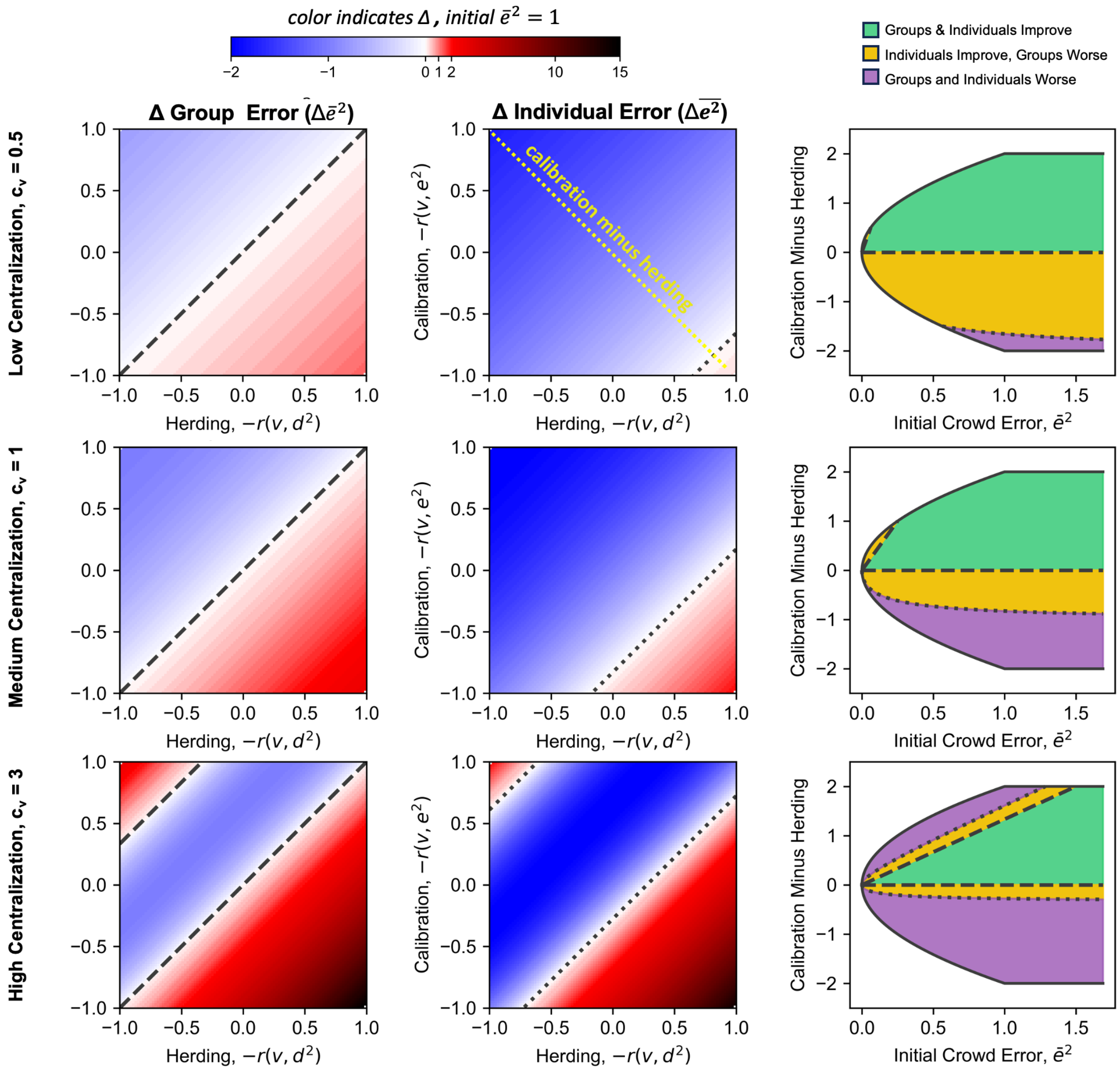

**Figure 1.** Change in error as a function of calibration and herding. The y-axis of the rightmost panels is equivalent to the upper-left-to-lower-right diagonal axis in the heatpams, and the heatmaps each represent a vertical slice of the rightmost figures. **LEFT/CENTER:** Change in error for groups (left) and individuals (center) as a function of calibration and herding. Under moderate centralization, individuals nearly always improve, even when groups get worse. Group accuracy always reduces when herding is greater than calibration, and generally improves when calibration is greater than herding. For highly centralized networks, very high calibration can paradoxically reduce group error as the group adopts "the wisdom of the few" rather than the wisdom of the crowd. Asymmetric color scale reflects the left-bounded value for the range -2<Δ<infinity. **RIGHT**: Change in error is determined by centralization (row), initial group error (x-axis), and the difference between calibration and herding (y-axis). Both groups and individuals generally improve (green area) when calibration is greater than herding. Individual accuracy can improve even when the group becomes less accurate (orange area). Both group and individual accuracy are reduced (purple) when herding is much stronger than calibration. **ALL:** $s_e$=1, $s_{e^2}$=1, $s_{d^2}$=1.

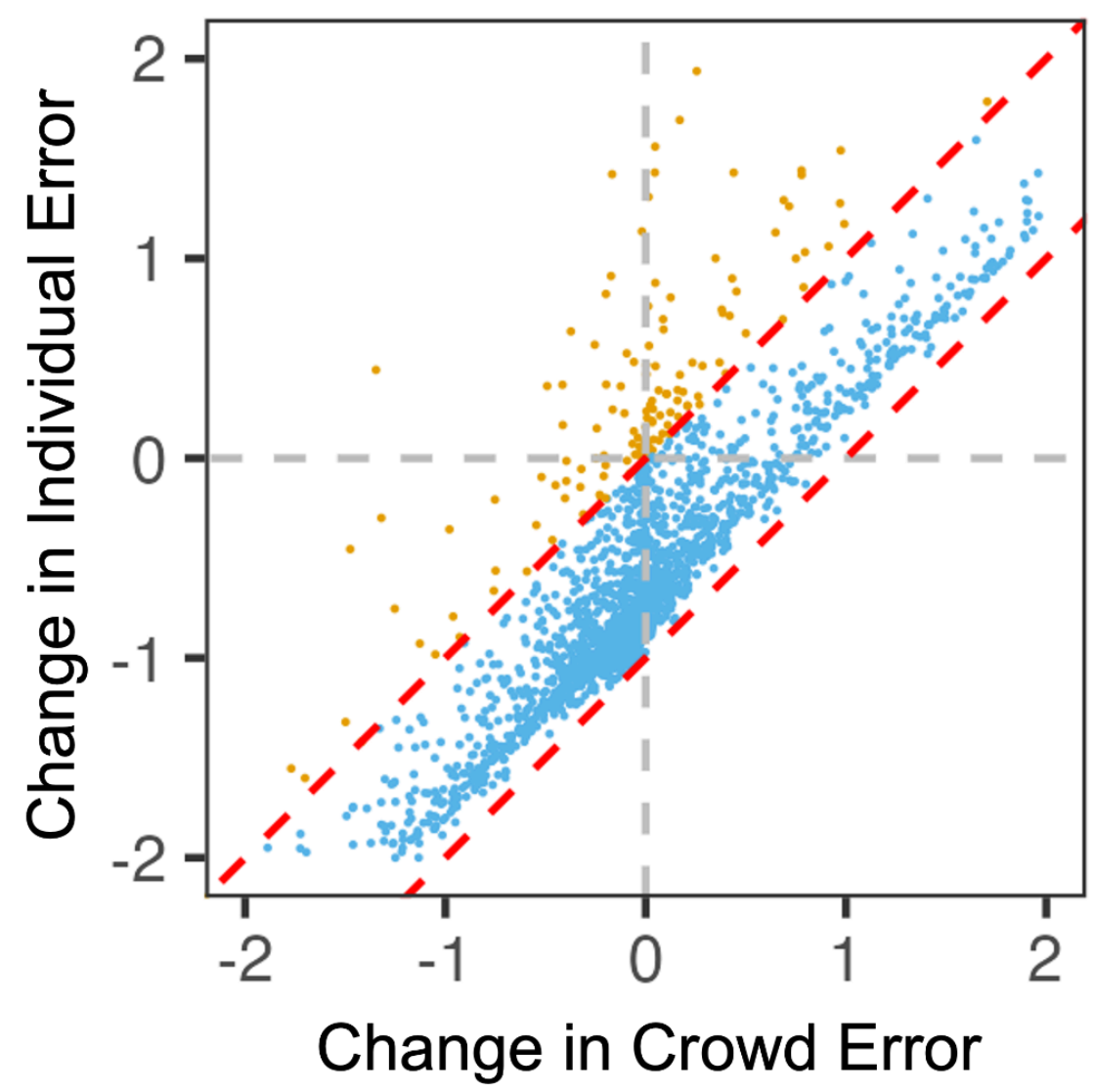

**Figure 2.** The relationship between group error and individual error. Each datapoint shows change in variance-standardised group and individual error for a single experimental trial. Blue points show trials where variance decreased, and group error improved less than individual error. Yellow points show trials where variance increased, an empirically rare condition where group error improves more than individual error. Red dashed diagonal lines show the boundaries Y=X and Y=X-1, with 90% of datapoints between the lines as expected. Grey dashed lines provide visual guides. Consistent with this equation and finite-time expectations, the regression slope=1 with an intercept -1<B<0 (see main text). To aid with visual representation this figure is "zoomed in" and shows only 85% of data. Appendix Fig. A2 shows a larger scale including 97% of data, illustrating the same relationship.

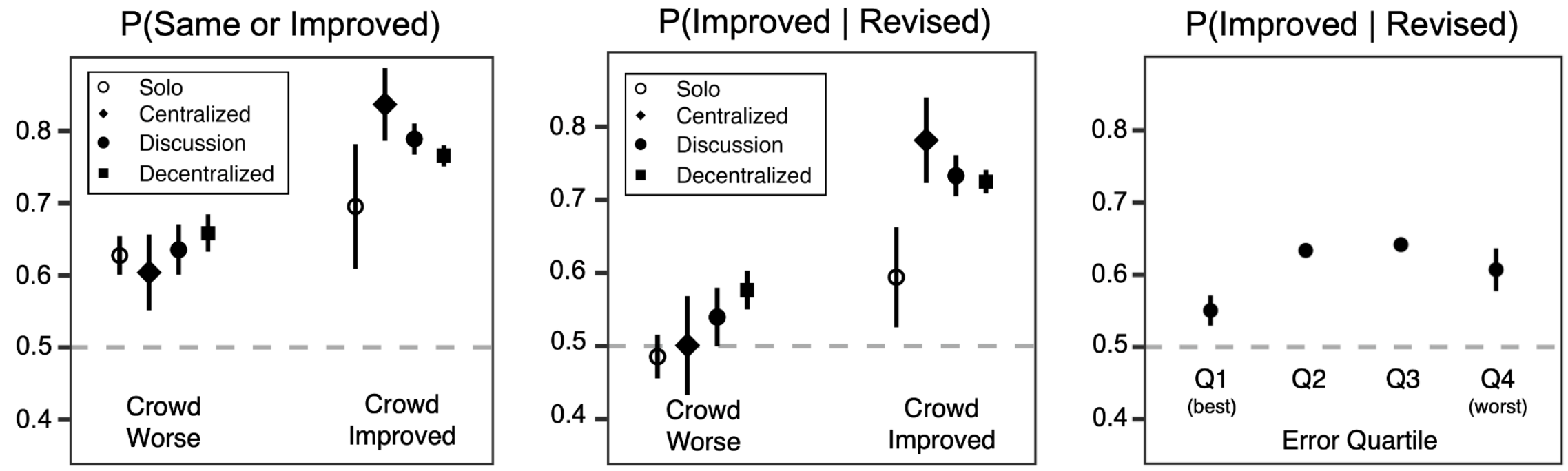

**Figure 3.** Probability of individuals benefiting from social influence with 95% bootstrapped confidence intervals. **LEFT AND MIDDLE:** The probability of improving as a function of group improvement, defined as simply not getting worse (middle) or improving conditional upon revision (left). **RIGHT:** Improvement as a function of an individual's relative accuracy, for social (non-solo) trials. Prior accuracy is measured only for non-focal questions to avoid regression to the mean (Q1 = top 25% most accurate, Q4 = bottom 25% least accurate*)*. Because error quartile is averaged across multiple questions, error quartiles do not necessarily contain 25% of the sample. Figure A3 shows results broken down by dataset.

## Appendix for

# Individuals, Crowds, and the Network Dynamics of Belief Accuracy

### A1. Changes in Individual and Group Error

**A1.1 Opinion and Error.** We assume that individuals have an initial opinion, $x_i$, that can be represented on some continuous numerical scale. We further assume that there is a value of the truth on this scale, $\theta$. Each individual has a bias that we define as

$$e_i = x_i - \theta. \quad \text{(A1)}$$

We define the *individual error* as the average of the squared bias

$$\overline{e^2} = \frac{\Sigma e_i^2}{n}, \quad \text{(A2)}$$

and the *group error* as the square of the average bias

$$\bar{e}^2 = \left(\frac{\Sigma e_i}{n}\right)^2. \quad \text{(A3)}$$

**A1.2 Relationship between Individuals and Groups.** The diversity prediction theorem (14) relates the average individual error and the squared error as

$$\overline{e^2} = \bar{e}^2 + s_e^2, \quad \text{(A4)}$$

where $s^2_e$ is the sample variance of errors. This relationship is equivalent to the definition of sample variance.

The diversity prediction theorem applied to initial opinions gives, $\overline{e^2} = \bar{e}^2 + s_e^2$, and final opinions, $\overline{e_k^2} = \bar{e}_k^2 + s_{e,k}^2$. Subtracting the relationship for final opinions by the relationship for initial opinions, we find that the change in average individual error is equal to the change in group error plus the change in opinion variance,

$$\Delta\overline{e^2} = \Delta\bar{e}^2 + \Delta s_e^2. \quad \text{(A5)}$$

Here, the change in average individual error is simply the final average individual errors minus the initial average individual errors $\Delta\overline{e^2} = \overline{e_k^2} - \overline{e^2}$. Similarly, the change in group error is the final minus the initial group error, $\Delta\bar{e}^2 = \bar{e}_k^2 - \bar{e}^2$. The change in sample variance is the final minus the initial sample variance $\Delta s_e^2 = s_{e,k}^2 - s_e^2$.

### A2. Experts

We consider a focal individual who has a level of accuracy that is described by the variance that their opinions tend to have around the truth,$\sigma_i^2$ We characterise the aggregated opinion of the group (excluding our focal individual) similarly, with a group accuracy that can be captured by the variance that the aggregated group opinion tends to have around the truth, $\sigma^2_{group}$ . We assume that the initial opinions of the individual and the group are independent.

Note that individuals, or the group, could also have some systematic bias such that their estimates are not centered on the truth. If this is known by the individual, then it can in principle be taken into account and opinions adjusted to remove the bias. If the bias is not known, then the uncertainty associated with the bias can essentially be considered a part of the variance.

We assume that an individual's updated opinion can be described as a weighted average of their own initial opinion and the initial group opinion, giving their own opinion a self-weight of $w_i$ and the group opinion a weight of $1 - w_i$.

In general, the variance of the weighted sum of independent random variables is $var(w_1A + w_2B) = w_1^2\sigma_1^2 + w_2^2\sigma_2^2$. Applying this identity to the updated individual opinion,

$$var(x_{i,k}) = w_i^2\sigma_i^2 + (1 - w_i)^2\sigma^2_{group} . \quad \text{(A6)}$$

The focal individual's accuracy is expected to improve through communication with the group when their initial opinion variance is greater than their updated variance, $\sigma_i^2 > var(x_{i,k})$. Substituting,

$$\sigma_i^2 > w_i^2\sigma_i^2 + (1 - w_i)^2\sigma^2_{group} . \quad \text{(A7)}$$

Rearranging

$$\sigma_i^2 > \frac{(1-w_i)^2}{1-w_i^2}\sigma^2_{group} . \quad \text{(A8)}$$

The denominator can be expressed as $1 - w_i^2 = (1 - w_i)(1 + w_i)$. Substituting and simplifying,

$$\sigma_i^2 > \frac{1-w_i}{1+w_i}\sigma^2_{group}. \quad \text{(A9)}$$

For most individuals, we would expect the group variance to be lower than the individual variance, due to the wisdom of the crowds effect (note that the group variance relates to the aggregated group opinion). In this case, the inequality is satisfied for any value of $w_i$, and the individual will benefit from listening to the group no matter what self-weight they have.

Experts may be more accurate than the group. In this case the expert can still benefit from listening to the group, as long as they do not give the group too much weight. Note that the right hand side can be made arbitrarily small by increasing the self-weight close to 1. This means that the condition can always be met with a non-zero group weight, for any values of individual and group variance. In practice, this means that an expert can always benefit by listening to the group, as long as they do not severely underestimate their own accuracy compared to the group.

Beyond conditions for improvement, differentiating Equation A6 shows that the self-weight that minimises the variance of the updated opinion is found through inverse-variance weighting,

$$w^*_i = \frac{\sigma^2_{group}}{\sigma^2_i + \sigma^2_{group}}. \quad \text{(A10)}$$

Which is always non-zero and positive, i.e. the optimal strategy for each individual is always to give some non-zero weight to the group.

**A3. Network Model**
In order to explore the conditions that determine how group and individual accuracy change, we make an assumption that opinions are updated under the DeGroot (31) model of social influence. At each timestep all individuals, $i$, revise their beliefs, $x_i$, by adopting a weighted mean of peer beliefs.

$$x_{i,t+1} = \sum_j w_{ij} x_{j,t} , \tag{A11}$$

where $w_{i,j}$ is the degree of one-step influence that agent *j* has on agent $i$.

**A3.1 Asymptotic Opinion and Influence.** Under mild assumptions (sufficient conditions are an aperiodic and strongly connected influence network), groups will eventually converge on a weighted mean of initial beliefs. This post-influence belief is weighted by the normalized leading eigenvector of the influence matrix (31), $v^{\infty}$, which can be interpreted as a vector of network centrality

$$x_{i,\infty} = \sum_j v^{\infty}_j x_j . \tag{A12}$$

**A3.2 One Step Opinion and Influence.** After one timestep, the group opinion is

$$\overline{x_1} = \frac{\sum_i x_{i,1}}{n} . \tag{A13}$$

Substituting in the DeGroot update rule (Equation A11),

$$\overline{x_1} = \frac{\sum_i \sum_j w_{ij} x_j}{n} . \tag{A14}$$

We can bring the summation over *i* inside the summation over *j*,

$$\overline{x_1} = \frac{\sum_j (\sum_i w_{ij}) x_j}{n} . \tag{A15}$$

We can recognise that $v^1_j = \frac{\sum_i w_{ij}}{n}$ is the mean out-weight that individual *j* has across all their peers, which is a measure of the one-step influence of individual *j*. We can write (switching the index of the outer summation from *j* to *i*),

$$\overline{x_1} = \sum_i v^1_i \, x_i . \tag{A16}$$

That is, the one-step group opinion is a weighted mean of the initial opinions, with the weights of each individual's opinion given by their one-step influence.

**A3.3. k-Step Opinion and Influence.** We can describe the influence weights, $w_{i,j}$, with an influence matrix. The influence matrix is stochastic, so that raising the matrix to the power of k gives entries $w^k_{ij}$, which describes the total influence that individual *i* has on individual *j* after *k* steps. (Note that the notation $w^k_{ij}$ does not refer to $w_{i,j}$ raised to the power of $k$, but instead the influence matrix raised to the power of $k$, and the $i,j$ entry of that matrix). After $k$ steps, the opinion of individual $i$ is a *k*-weighted mean of the initial opinions of peer opinions,

$$x_{i,k} = \sum_j w^k_{ij} x_j . \tag{A17}$$

This is a similar form to the one step DeGroot update rule. We can follow the logic of the derivation for one-step group opinion and conclude that the k-step group opinion is given by

$$\overline{x_k} = \sum_i v^k_i \, x_i , \tag{A18}$$

where $v^k$ is the k-step influence vector, with entries $v^k_i$ given by the k-step mean outweight that individual *i* has on their peers.

Note that the k-step influence generalises to the asymptotic case. In the limit $k \rightarrow \infty$, the k-step influence vector converges to the normalised leading eigenvector of the influence matrix, $v^{\infty}$.

In the main paper, and from here in the Appendix, we refer generally to an influence vector *v*, which can represent any *k*-step influence vector without loss of generality.

**A3.4. Change in Opinion**

The empirical covariance between two variables is defined as $cov(A,B) = \overline{AB} - \overline{A}\,\overline{B}$, where $\overline{A}$ and $\overline{B}$ are the empirical means of each variable and $\overline{AB}$ is the empirical mean of the product of the two variables. Considering an influence vector and the initial opinions, we can use this identity to write

$$cov(v\ ,\ x) = \frac{\sum vx}{n} - \frac{\sum v}{n}\frac{\sum x}{n}\ . \qquad \text{(A19)}$$

By definition, the influence vector is normalised such that $\sum v = 1$. Substituting and rearranging, we can write

$$\sum vx = n\, cov(v,\ x) + \overline{x}\ . \qquad \text{(A20)}$$

The summation is equivalent to the k-step group opinion (Equation A18). Substituting,

$$\overline{x_k} = n\, cov(v,\ x) + \overline{x}\ . \qquad \text{(A21)}$$

The change in group opinion after k-steps is then

$$\overline{x_k} - \overline{x} = n\, cov(v,\ x)\ . \qquad \text{(A22)}$$

The covariance can instead be written in terms of correlation, $cov(v,\ x) = s_v s_x r(v,x)$, such that

$$\overline{x_k} - \overline{x} = n\, s_v\, s_x\, r(v,x). \qquad \text{(A23)}$$

The coefficient of variation of a variable is the standard deviation divided by the mean, $c_A = \frac{s_A}{\overline{A}}$. in the case of the influence vector the mean influence is $\overline{v} = \frac{1}{n}$, so that the coefficient of variation of influence is $c_v = ns_v$. Substituting and rearranging we find the relationship for the standardised change in k-step opinion

$$\frac{\overline{x_k} - \overline{x}}{s_x} = c_v\, r(v,x). \qquad \text{(A24)}$$

We stress that this is not an approximation or an expectation, this is the precise change in group opinion after k-steps of DeGroot updating. Note that the relationship is scale free (does not depend on *n*).

We therefore find that the standardized *k*-step change in group opinion is scale free and a linear product of i) a measure of *k*-step influence centralisation, $c_v$ and ii) the correlation between influence and initial opinions, $r(v,x)$.

### A3.5. Bounds on the coefficient of variation

The coefficient of variation of influence is defined as $c_v = n s_v$. The empirical variance of influence is by definition $s_v^2 = \frac{\sum_i (v_i - \bar{v})^2}{n}$. Substituting,

$$c_v^2 = n \sum_i (v_i - \frac{1}{n})^2. \quad \text{(A25)}$$

The minimal influence centralisation is given by a fully decentralised network where all individuals have influence $v_i = \frac{1}{n}$, in which case $c_v = 0$.

The maximal influence centralisation is a fully centralised network where one individual has influence 1, and all other individuals have influence 0. In this case

$$c_v^2 = n\left((1 - \frac{1}{n})^2 + (n - 1)(\frac{1}{n})^2\right), \quad \text{(A26)}$$

expanding and simplifying,

$$c_v^2 = n - 1. \quad \text{(A27)}$$

The coefficient of variation of influence is therefore bounded such that $0 \le c_v \le \sqrt{n - 1}$.

Note that the upper bound depends on the number of individuals, *n*. The coefficient of variation of influence is scale free such that if we change the size of a network while maintaining the same influence distribution (and renormalising) then the coefficient of variation stays the same, unlike the standard deviation of influence. Larger networks can have larger coefficients of variation because larger networks can have more centralised influence distributions. For example, a network of size 3 with maximum centralisation will have an influence distribution of the form (1,0,0). If we double the network size to 6 but maintain the same distribution then we have (½, ½, 0,0,0,0), and the coefficient of variation in this case is equivalent in the two networks. However, the network with size 6 can be more centralised with a distribution (1,0,0,0,0,0).

### A3.6. Bounds on Change In Opinion

We can also write down bounds on the change in opinion. Considering Equation A24, the correlation $r(v, x)$ is bounded between -1 and 1, so that the change in opinion is bounded by

$$-c_v \le \frac{\bar{x}_k - \bar{x}}{s_x} \le c_v. \quad \text{(A28)}$$

Considering the bounds on the coefficient of variation, we find the standardised change in group opinion is bounded by the number of individuals,

$$-\sqrt{n - 1} \le \frac{\bar{x}_k - \bar{x}}{s_x} \le \sqrt{n - 1}. \quad \text{(A29)}$$

### A3.7. DeGroot Rule For Bias

Given the truth, θ, we convert opinions to biases through a linear transformation, $e_{i,t} = x_{i,t} - \theta$. Through substitution into the DeGroot updating rule for opinions (Equation A11) we find the DeGroot rule for bias,

$$e_{i,t+1} = \sum_j w_{ij} e_{j,t}. \quad \text{(A30)}$$

We note that the truth cancels on both the left and right hand sides so that this is of the same form as Equation A11, and that the analysis above (Equations A12-A30) are also in the same form when considering biases instead of opinions.

**A3.8. Asymptotic Change in Error (Squared Bias)**
Earlier, we found that the change in group bias (or equivalently change in group opinion) can be written

$$\frac{\bar{e}_k - \bar{e}}{s_e} = c_v r(v, e) . \tag{A31}$$

Rearranging, we can write the k-step group bias as

$$\bar{e}_k = s_e c_v r(v, e) + \bar{e} . \tag{A32}$$

The change in group opinion is defined as $\Delta\bar{e}^{\,2} = \bar{e}_k^{\,2} - \bar{e}^{\,2}$. Substituting the k-step group bias,

$$\Delta\bar{e}^{\,2} = (s_e c_v r(v, e) + \bar{e})^2 - \bar{e}^2 . \tag{A33}$$

Expanding, simplifying, rearranging, we find the standardised change in group error

$$\frac{\Delta\bar{e}^{\,2}}{s_e^{\,2}} = c_v^{\,2} r(v, e)^2 + 2\,\frac{\bar{e}}{s_e} c_v r(v, e) , \tag{A34}$$

where $z = \frac{\bar{e}}{s_e}$ is the standardised initial group bias. Substituting this relationship into Equation A5, relating changes in group error and changes in individual error, we find the standardised change in individual error,

$$\frac{\Delta\overline{e^2}}{s_e^{\,2}} = c_v^{\,2} r(v, e)^2 + 2\,\frac{\bar{e}}{s_e} c_v r(v, e) + \frac{\Delta s_e^{\,2}}{s_e^{\,2}} . \tag{A35}$$

**A3.9 Group Opinion Beta**

The group opinion beta, β, provides a compressed way to represent changes in group opinion and error. In the main paper we define β as a normalised measurement of the final group bias as a fraction of the initial group bias

$$\beta = \frac{\bar{e}_k}{\bar{e}} . \tag{A36}$$

We can write change in group bias, $\Delta\bar{e} = \bar{e}_k - \bar{e}$, through substituting $\bar{e}_k = \beta\,\bar{e}$,

$$\Delta\bar{e} = \bar{e}\,(\beta - 1) . \tag{A37}$$

We can write the change in group error, $\Delta\bar{e}^2 = \bar{e}_k^2 - \bar{e}^2$, through substituting $\bar{e}_k^{\,2} = \beta^2\,\bar{e}^2$,

$$\Delta\bar{e}^2 = \bar{e}^{\,2}(\beta^2 - 1) . \tag{A38}$$

Through substituting $\bar{e}_k = \beta\,\bar{e}$ into the change in opinion/bias relationship (Equation A24), we find that

$$\beta = \frac{c_v r(v,e)}{z} + 1 . \tag{A39}$$

**A3.10. Decomposition Into Calibration and Herding**

We found analytic expressions for the changes in group error and changes in individual error. However they require knowledge of the direction of the group bias, *z*, in order to predict how a group (and individuals within the group) will change in accuracy. We derive alternative analytical representations that are independent of the direction of the initial group bias.

We define the "calibration" of the group, $C = -r(v, e^2)$. A highly calibrated group is one where influence aligns with accuracy. We also define the "herding" of the group, $H = -r(v, d^2)$, where $d^2$ is a vector of squared deviances such that $d_i = x_i - \bar{x}$. Bias is a linear transformation of opinions, so we can also write $d_i = e_i - \bar{e}$.

We can write down the covariance related to herding,

$$cov(v, d^2) = \overline{vd^2} - \bar{v}\,\overline{d^2}. \quad \text{(A40)}$$

Substituting in the definition of deviance,

$$cov(v, d^2) = \overline{\left(v\,(e - \bar{e})^2\right)} - \bar{v}\,\overline{\left((e - \bar{e})^2\right)}. \quad \text{(A41)}$$

Expanding the terms

$$cov(v, d^2) = \overline{\left(v\,e^2 - 2ve\,\bar{e} + v\,\bar{e}^2\right)} - \bar{v}\,\overline{\left(e^2 - 2e\,\bar{e} + \bar{e}^2\right)}. \quad \text{(A42)}$$

Taking means of each term and cancelling like terms,

$$cov(v, d^2) = \overline{ve^2} - 2\,\overline{ve}\,\bar{e} - \bar{v}\,\overline{e^2} + 2\,\bar{v}\,\bar{e}^2. \quad \text{(A43)}$$

We now do the same with the covariance related to calibration,

$$cov(v, e^2) = \overline{ve^2} - \bar{v}\,\overline{e^2}. \quad \text{(A44)}$$

We can combine these equations to find

$$cov(v, d^2) = cov(v, e^2) - 2\,\overline{ve}\,\bar{e} + 2\,\bar{v}\,\bar{e}^2. \quad \text{(A45)}$$

Factorising,

$$cov(v, d^2) = cov(v, e^2) - 2\,\bar{e}\,(\,\overline{ve} - \bar{v}\,\bar{e}\,). \quad \text{(A46)}$$

Substituting in the covariance relation $cov(v, e) = \overline{ve} - \bar{v}\,\bar{e}$,

$$cov(v, d^2) = cov(v, e^2) - 2\,\bar{e}\,cov(v, e). \quad \text{(A47)}$$

For each covariance term, we substitute the correlation relation $cov(a, b) = s_a s_b r(a, b)$,

$$s_{d^2} s_v r(v, d^2) = s_{e^2} s_v r(v, e^2) - 2\bar{e}\, s_e s_v r(v, e), \quad \text{(A48)}$$

where $s_{e^2}$ is the standard deviation in initial errors (squared biases), and $s_{d^2}$ is the standard deviation in initial squared deviances. Cancelling $s_v$ and rearranging,

$$r(v, e) = \frac{1}{2\bar{e}\, s_e}\left(s_{e^2} r(v, e^2) - s_{d^2} r(v, d^2)\right). \quad \text{(A49)}$$

This identity relates calibration, herding, the bias-influence correlation and the initial group bias.

We can substitute herding as H = -r(v,d2) and calibration as C = -r(v,e2),

$$r(v, e) \;=\; \frac{1}{2\bar{e}\, s_e}\left(s_d^2 H \;-\; s_e^2 C\right).$$

Substituting this into Equation A34, we can write the change in group error in terms of calibration and herding,

$$\Delta\bar{e}^{\,2} = \frac{c_v^{\,2}}{4\,\bar{e}^2}\left(s_d^2 H \;-\; s_e^2 C\right)^2 \;+ c_v\left(s_d^2 H \;-\; s_e^2 C\right)\;. \tag{A50}$$

Similarly, the change in individual error (substituting into Equation A35) is

$$\Delta\overline{e^{2}} = \frac{c_v^{\,2}}{4\,\bar{e}^2}\left(s_d^2 H \;-\; s_e^2 C\right)^2 \;+ c_v\left(s_d^2 H \;-\; s_e^2 C\right) \;\;+\; \Delta s_e^{\,2}\,. \tag{A51}$$

And the group Beta (substituting into Equation A39) is

$$\beta \;=\; \frac{c_v}{2\bar{e}^2}\left(s_d^2 H \;-\; s_e^2 C\right) \;+\; 1. \tag{A52}$$

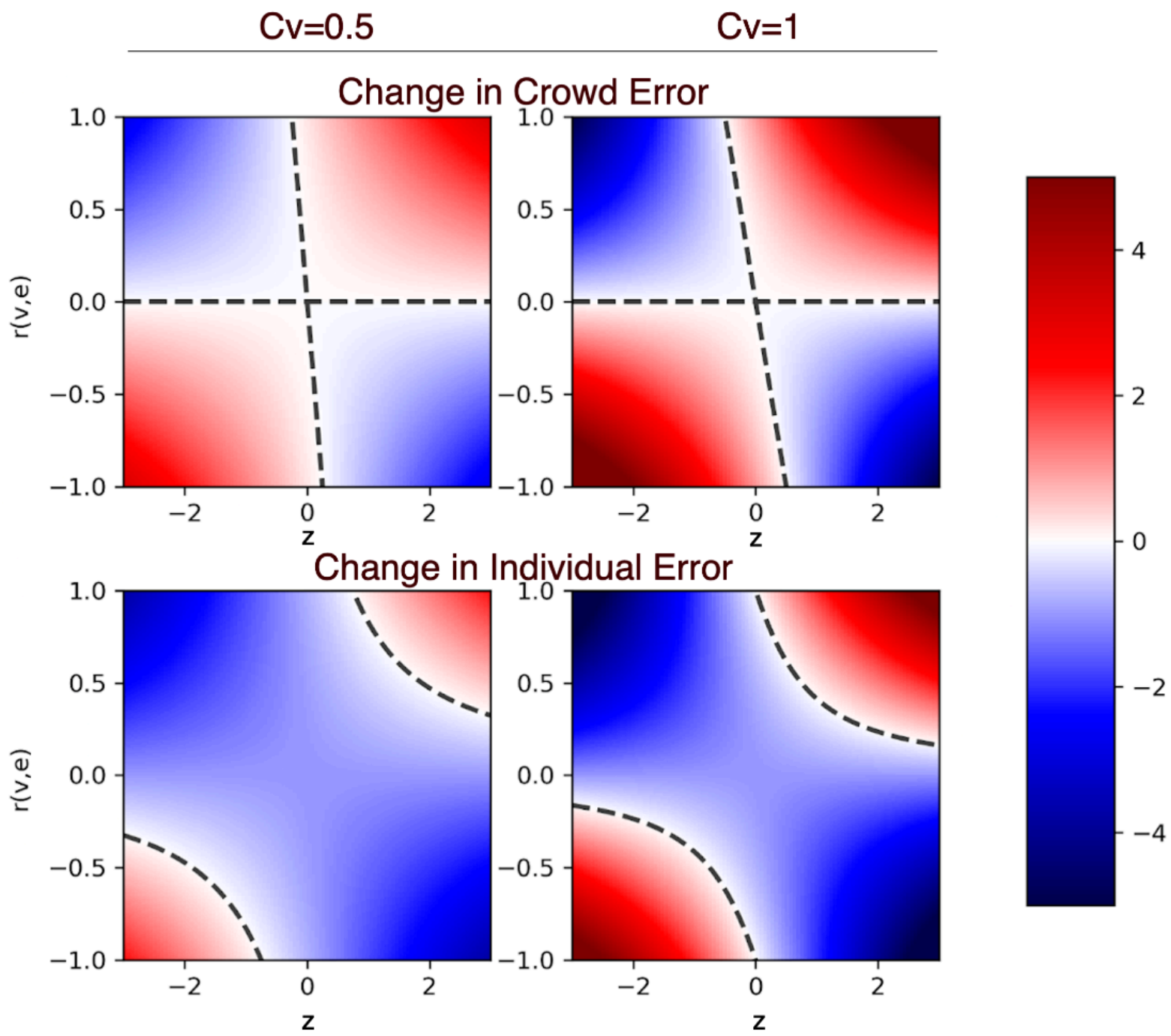

**Figure A1.** The change in group error (top) and individual error (bottom) with increasing influence centralization ($c_v$, left to right). as a function of initial group bias, *z*, and correlation between influence and initial bias, *r(v,e)*. Negative changes in error (blue) indicate an improvement in accuracy. Dashed lines show the boundaries of improvement in accuracy.

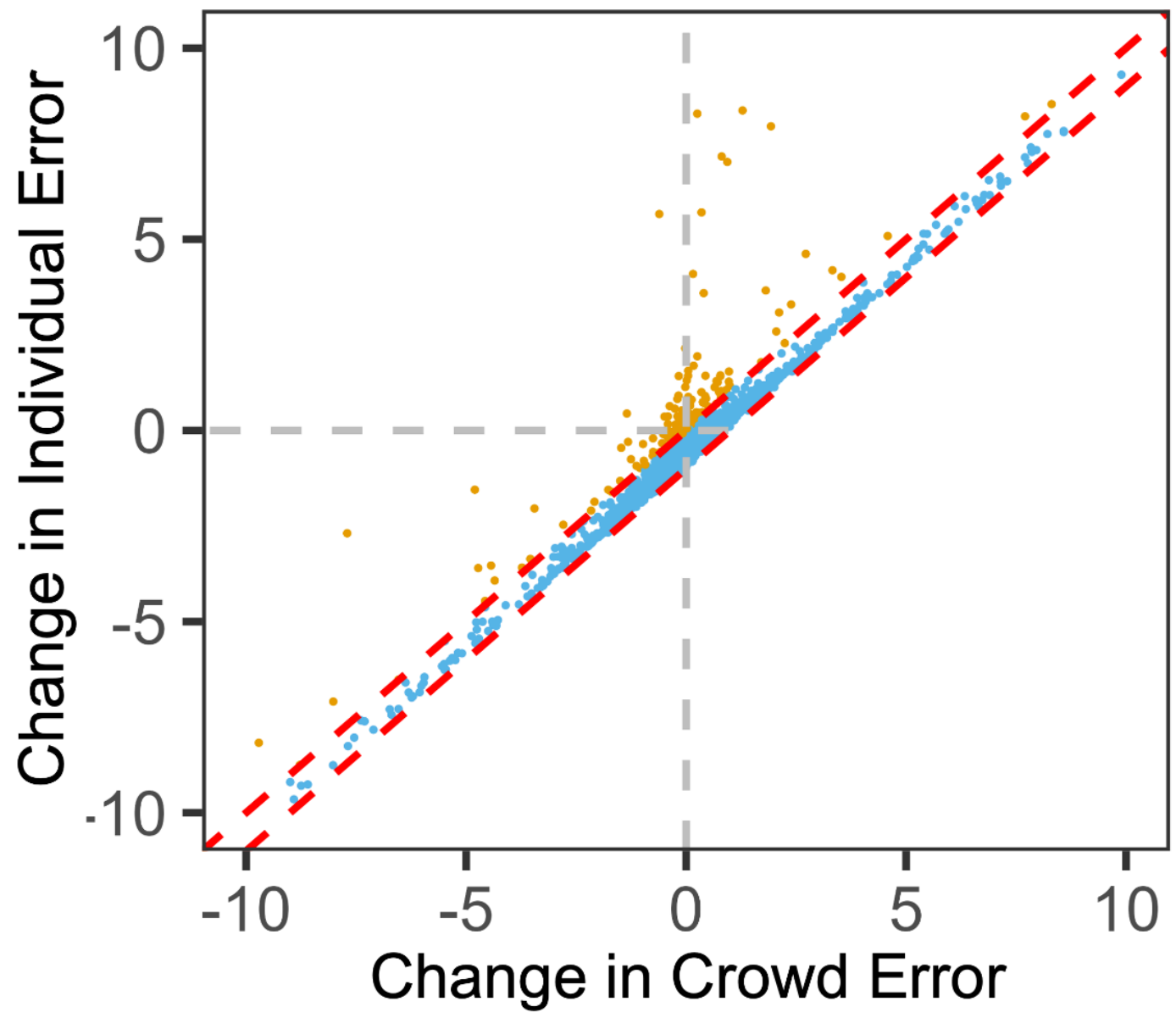

**Figure A2.** Figure 2, but "zoomed out" to show 97% of datapoints.

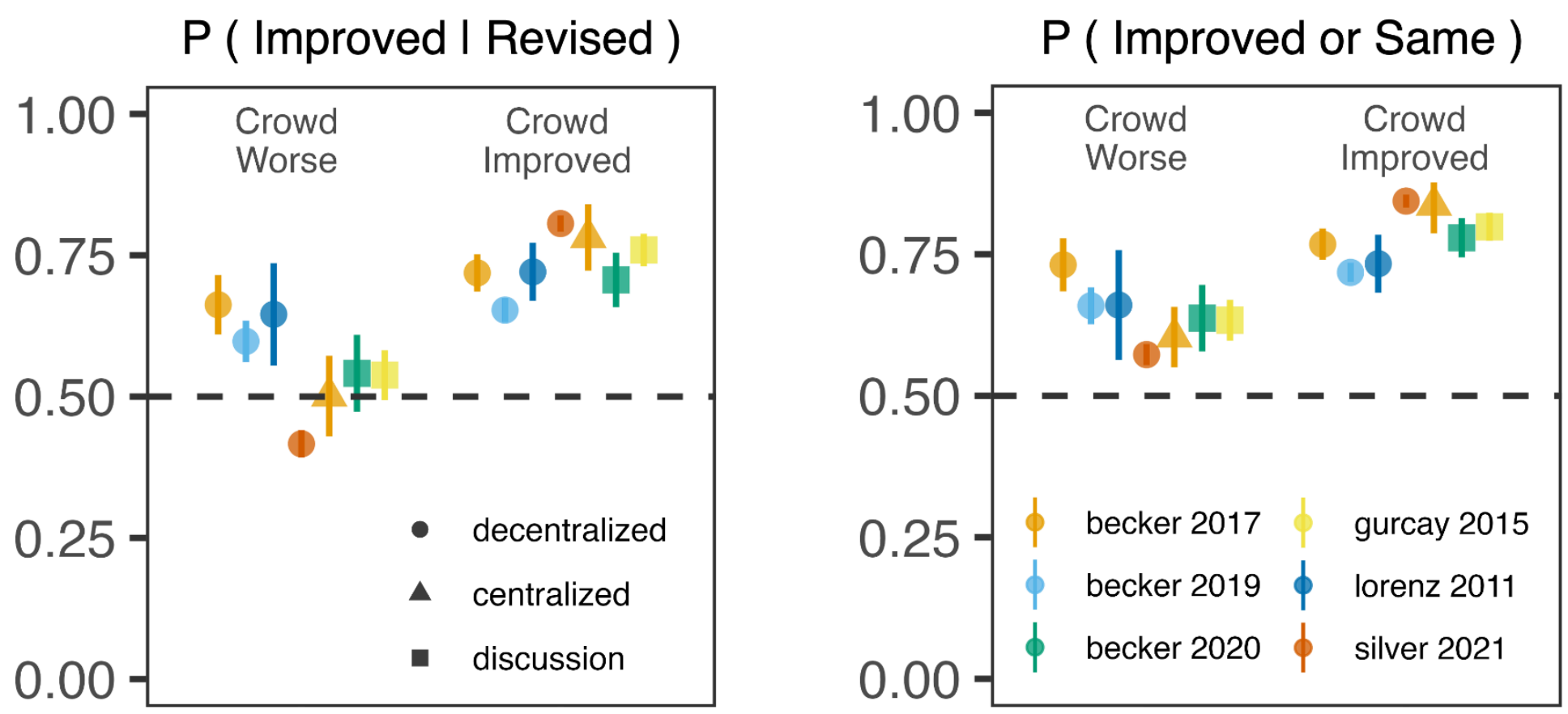

**Figure A3.** Figure 3, broken down by dataset. Color chart in right panel and shape chart in left panel indicate data types for both panels.

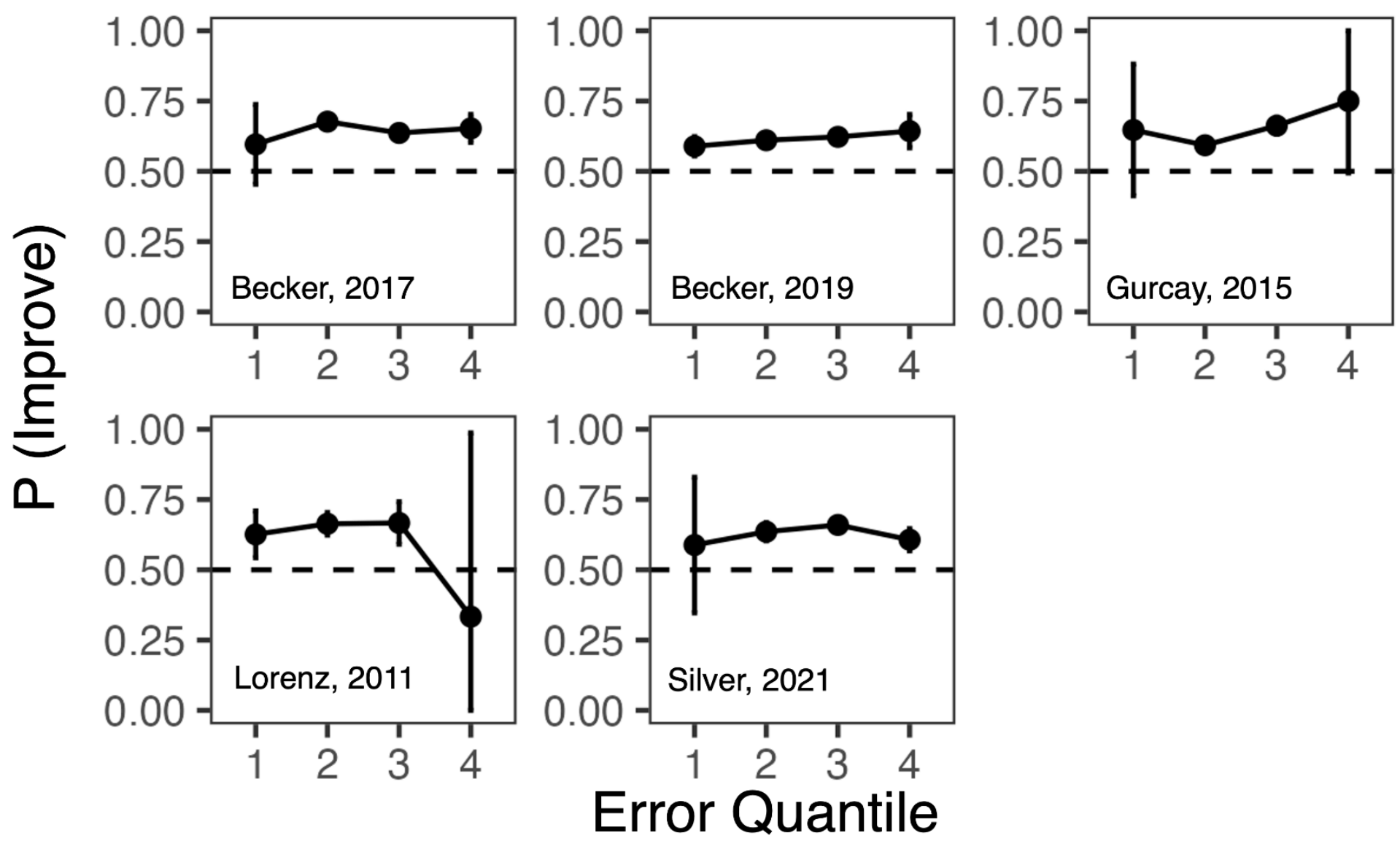

**Fig A4.** Fig 4, broken down by dataset.